\documentclass[preprint]{aastex61}

\usepackage{multirow}
\usepackage{booktabs}
\usepackage{rotating}

\hypersetup{linkcolor=red,citecolor=blue,filecolor=cyan,urlcolor=magenta}
\newcommand{\wise}{{\textit{WISE~}}}
\newcommand{\wone}{{\textit W1~}}
\newcommand{\wtwo}{{\textit W2~}}
\newcommand{\wthree}{{\textit W3~}}
\newcommand{\wfour}{{\textit W4~}}
\newcommand{\hst}{{\it HST~}}
\newcommand{\herschel}{{\textit{Herschel~}}}
\newcommand{\um}{$\mu m$~}

\newcommand{\ser}{S\'ersic~}
\newcommand{\galfit}{{\sf GALFIT~}}
\newcommand{\sedfit}{{\sf SED3FIT~}}

\shorttitle{The SED of Hot DOG W2246$-$0526}
\shortauthors{Fan et al.}
\begin{document}

\title{The spectral energy distribution  of the hyperluminous, hot dust-obscured galaxy W2246$-$0526}

\correspondingauthor{Lulu Fan}
\email{llfan@sdu.edu.cn}

\author{Lulu Fan}
\affil{Shandong Provincial Key Lab of Optical Astronomy and Solar-Terrestrial Environment, Institute of Space Science, Shandong University, Weihai, 264209, China}

\author{Ying Gao}
\affiliation{Shandong Provincial Key Lab of Optical Astronomy and Solar-Terrestrial Environment, Institute of Space Science, Shandong University, Weihai, 264209, China}

\author{Kirsten K. Knudsen}
\affiliation{Department of Space, Earth and Environment, Chalmers University of Technology, Onsala Space Observatory, SE-439 92 Onsala, Sweden}

\author{Xinwen Shu}
\affiliation{Department of Physics, Anhui Normal University, Wuhu, Anhui, 241000, China}

%% Note that the \and command from previous versions of AASTeX is now
%% depreciated in this version as it is no longer necessary. AASTeX
%% automatically takes care of all commas and "and"s between authors names.

%% AASTeX 6.1 has the new \collaboration and \nocollaboration commands to
%% provide the collaboration status of a group of authors. These commands
%% can be used either before or after the list of corresponding authors. The
%% argument for \collaboration is the collaboration identifier. Authors are
%% encouraged to surround collaboration identifiers with ()s. The
%% \nocollaboration command takes no argument and exists to indicate that
%% the nearby authors are not part of surrounding collaborations.

%% Mark off the abstract in the ``abstract'' environment.
\begin{abstract}

Hot dust-obscured galaxies (Hot DOGs) are a luminous, dust-obscured population recently discovered in the \wise All-Sky survey. Multiwavelength follow-up observations suggest that they are mainly powered by accreting supermassive black holes (SMBHs), lying in dense environments, and being in the transition phase between extreme starburst and UV-bright quasars. Therefore, they are good candidates for studying the interplay between SMBHs, star formation and environment. W2246$-$0526 (thereafter, W2246), a Hot DOG at $z\sim4.6$, has been taken as the most luminous galaxy known in the Universe. Revealed by the multiwavelength images, the previous \herschel SPIRE photometry of W2246 is contaminated by a foreground galaxy (W2246f), resulting in an overestimation of its total IR luminosity by a factor of about 2. We perform the rest-frame UV/optical-to-far-IR spectral energy distribution (SED) analysis with \sedfit and re-estimate its physical properties. The derived stellar mass $M_\star = 4.3\times10^{11}~M_\odot$ makes it be among the most massive galaxies with spectroscopic redshift $z>4.5$. Its structure is extremely compact and requires an effective mechanism to puff-up. Most of ($>95\%$) its IR luminosity is from AGN torus emission, revealing the rapid growth of the central SMBH. We also predict that W2246 may have a significant molecular gas reservoir based on the dust mass estimation.  

\end{abstract}

%% Keywords should appear after the \end{abstract} command.
%% See the online documentation for the full list of available subject
%% keywords and the rules for their use.


%% From the front matter, we move on to the body of the paper.
%% Sections are demarcated by \section and \subsection, respectively.
%% Observe the use of the LaTeX \label
%% command after the \subsection to give a symbolic KEY to the
%% subsection for cross-referencing in a \ref command.
%% You can use LaTeX's \ref and \label commands to keep track of
%% cross-references to sections, equations, tables, and figures.
%% That way, if you change the order of any elements, LaTeX will
%% automatically renumber them.

%% We recommend that authors also use the natbib \citep
%% and \citet commands to identify citations.  The citations are
%% tied to the reference list via symbolic KEYs. The KEY corresponds
%% to the KEY in the \bibitem in the reference list below.

\section{Introduction}

One of the primary science objectives for {\it Wide-field Infrared Survey Explorer} \citep[{\it WISE};][]{wright2010} all-sky survey is to identify the most luminous ultraluminous infrared galaxies (ULIRGs) in the Universe. With a so-called {\it W1W2}-dropout color-selected method \citep{eisenhardt2012,wu2012}, a new population of luminous, dust-obscured galaxies \citep[designated as Hot, Dust-Obscured Galaxies, for short Hot DOGs, by][]{wu2012} has been successfully discovered.  Several works have suggested that Hot DOGs, mainly powered by accreting supermassive black holes (SMBHs), may represent a key transition phase during the evolution of massive galaxies, linking starbursts and luminous unobscured quasars \citep{wu2012,bridge2013,diaz-santos2016,fan2016a,fan2016b,wu2018}. 

Among those Hot DOGs with spectroscopic redshift and far-infrared photometry, W2246$-0526$ (thereafter, W2246) is the most distant one at redshift $z_{opt}=4.593$ derived from UV/optical emission lines \citep{wu2012,tsai2015}. With ALMA [CII] observations of W2246, \citet{diaz-santos2016} measured its redshift  at $z_{\rm [CII]}$=4.601 which shows [CII] line having a significant redshift compared with UV/optical emission lines. The previous works used multiwavelength fit to its SED and obtained its total IR luminosity $L_{\rm IR} = 2.2-3.4\times10^{14}L_\odot$ \citep{tsai2015,fan2016b}. Given the corresponding bolometric luminosity $L_{\rm bol} = 3.5-4.8 \times 10^{14} L_\odot$, W2246 had been taken as the most luminous galaxy known in the Universe \citep{diaz-santos2016}. 

\begin{figure*}
\centering
\includegraphics[width=\hsize]{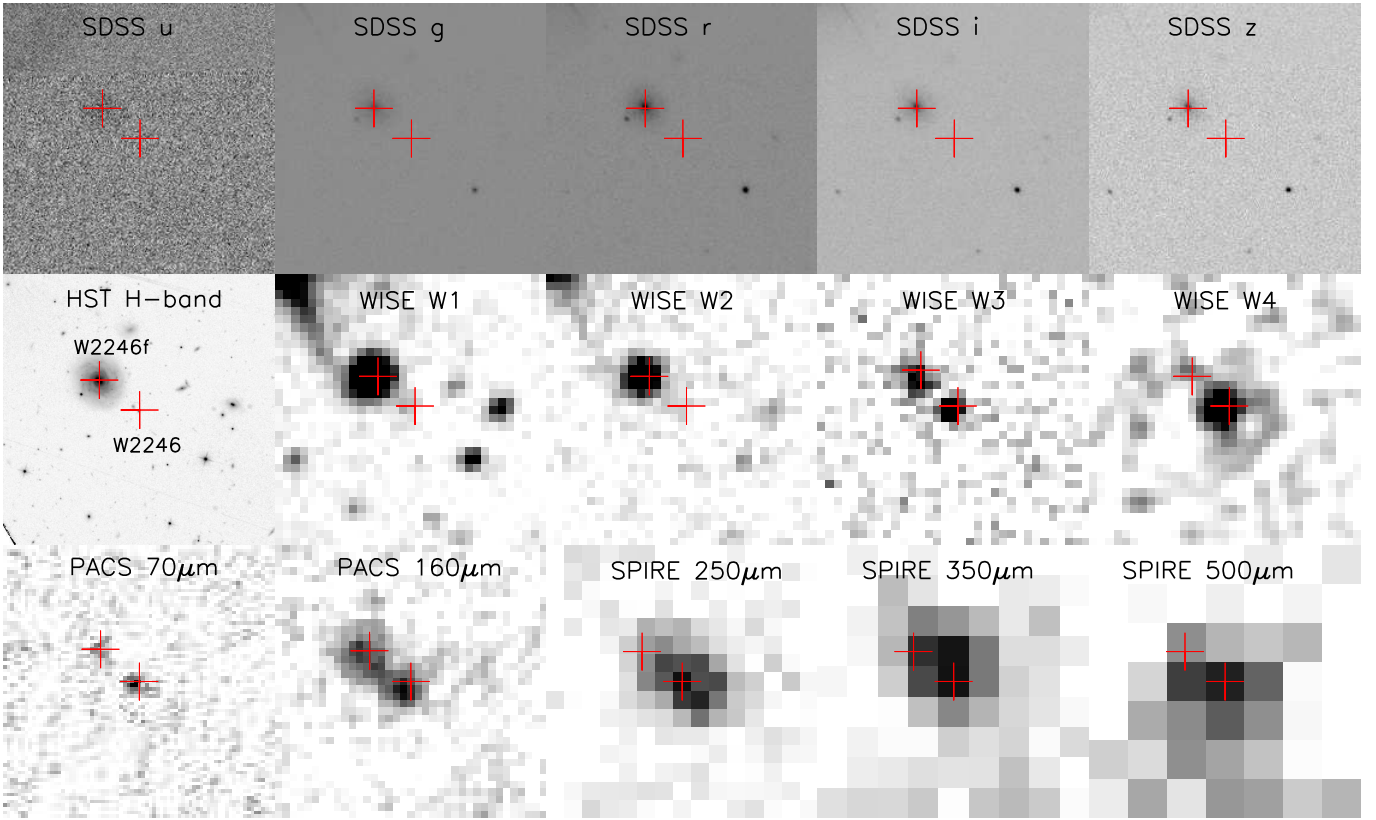}
\caption{Multiwavelength images of W2246 and its nearby foreground galaxy, W2246f. Top panels: SDSS \textit{u-,g-,r-,i-} and \textit{z-}band images. Middle panels: \hst H-band, \wise {\it W1}, {\it W2}, \wthree and \wfour maps, taken from unWISE images \citep{lang2014}. Bottom panels: from left to right, \herschel maps in PACS 70  and 160~$\mu m$, SPIRE 250, 350, and 500~$\mu m$, respectively. These maps are 1.5$\times$1.5 square arcmin in size and centered on W2246. Two crosses mark the positions of W2246 and W2246f based on \hst H-band image. }
\label{fig:stamps}
\end{figure*}

Recently, we noted that the IR luminosity of W2246 was likely overestimated due to the contamination of a foreground galaxy to \herschel SPIRE photometry. In Figure \ref{fig:stamps}, we show the multiwavelength images of W2246 and the nearby foreground galaxy, W2246f, which is about 16$''$ away to the northeast of W2246. Due to the poor resolution of \herschel SPIRE 250, 350, and 500~\um bands, it is clear that \herschel SPIRE photometry of W2246 is significantly affected by the contamination of W2246f. We re-measure the \herschel SPIRE flux with point spread function (PSF) fitting. With the updated flux estimations, we fit the rest-frame UV/optical-to-far-IR SED of W2246 with \sedfit code \citep{berta2013} and obtain the key physical properties. Throughout this work we assume a standard, flat ${\rm \Lambda}$CDM cosmology \citep[see][]{komatsu2011}, with $H_0 = 70$ km~s$^{-1}$, $\Omega_M = 0.3$, and $\Omega_\Lambda = 0.7$.

%(22h46m08.39s; -05d26m24.6s; J2000.0)
%------------------------------------------------------
%
\begin{figure}
\centering
\includegraphics[width=0.8\hsize]{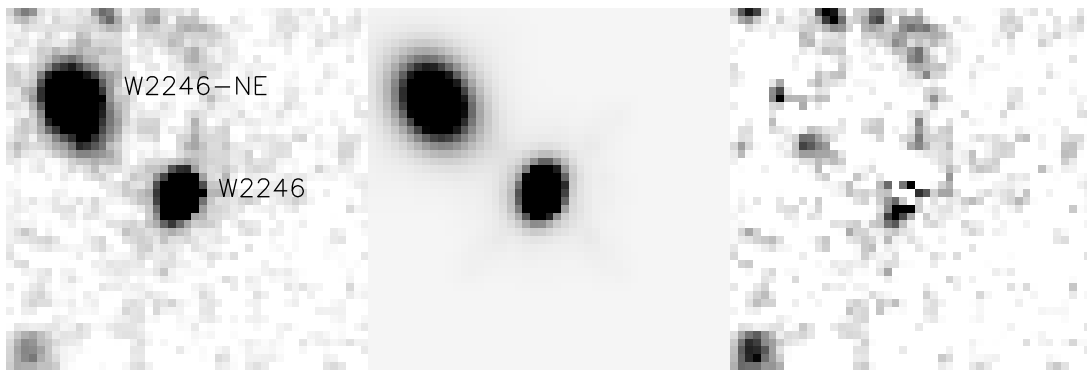}
\caption{ Zoom-in \hst H-band image ($6''\times 6''$), including W2246 in the center and its northeast companion (W2246-NE), is presented in the left panel. We use the \galfit package to fit the surface brightness profiles of W2246 and W2246-NE, simultaneously. For W2246, we adopt a \ser + PSF model to represent the host galaxy and AGN component, respectively. For W2246-NE, we only assume a \ser model. The model and residual images of W2246 and W2246-NE have been shown in the middle and right panels, respectively.         
         }
\label{fig:hstw2246}
\end{figure}
%-------
%--------------------------------------------------------------------
\section{Data} \label{sec:data}

In order to construct the rest-frame UV/optical-to-far-IR SEDs of W2246 and W2246f, we compile the available multiwavelength data in the literature. In Figure \ref{fig:stamps}, we show the multiwavelength images centering on W2246, including five SDSS bands (\textit{u, g, r, i} and \textit{z}), \hst H-band, four \wise bands ({\it W1}, {\it W2}, {\it W3} and {\it W4}),  two \herschel  PACS bands \citep[70  and 160~$\mu m$,][]{poglitsch2010} and three SPIRE bands \citep[250, 350, and 500~$\mu m$,][]{griffin2010}. Two red crosses mark the positions of W2246 (22h46m07.5s,$-05$d26m35.3s) and W2246f (22h46m08.3s,$-05$d26m24.5s) based on the \hst H-band image. 

Five-SDSS-band photometry and photometric redshift of W2246f have been retrieved from Sloan Digital Sky Survey SkyServer\footnote{http://skyserver.sdss.org/}. The photometric redshift of W2246f ($z_{\rm ph} = 0.047$) suggests that it is a foreground galaxy. Instead of retrieving the \wise {\it W1}, {\it W2}, {\it W3} and {\it W4} photometry from the \wise ALLWISE Data Release \citep{cutri2013}, we do the aperture photometry of both W2246 and W2246f based on the unblurred coadded \wise images\footnote{https://unwise.me} \citep[unWISE,][]{lang2014}. The photometry errors have been estimated based on the inverse variance images. 
%We use zero points of 306.68, 170.66, 29.04 and 8.284 Jy to convert {\it W1}, {\it W2}, {\it W3} and {\it W4} Vega magnitudes to flux densities, respectively \citep{wright2010}. 

W2246f has the 2MASS photometry. We retrieve its near-IR \textit{J, H} and $K_S$ flux densities from NED\footnote{https://ned.ipac.caltech.edu}. In Figure \ref{fig:hstw2246}, we show a zoom-in $6''\times 6''$ \hst H-band image. High-resolution \hst H-band image reveals two components: W2246 and its northeast companion (W2246-NE: 22h46m07.6s,$-05$d26m33.7s). We derive their photometry by using the \galfit package \citep{peng2002,peng2010} to fit the surface brightness profiles of W2246 and W2246-NE, simultaneously. For W2246, we adopt a \ser + PSF model to represent the host galaxy and AGN component, respectively. For W2246-NE, we only assume a \ser model. The model PSF of \hst H-band has been constructed by using the {\sf TinyTim} package \citep{krist2011}. {\bf The \ser index $n$ has been set as a free parameter. The best-fit models suggest that both W2246 and W2246-NE have disk-like structure, with $n= 0.7$ and $0.8$, respectively.} The model and residual images of W2246 and W2246-NE have been shown in Figure \ref{fig:hstw2246}. {\bf The remaining flux on the residual image contributes less than 3\% of the total flux of W2246. The pattern of the residual image of W2246 suggests that the system is possibly not relaxed yet after a recent merger.} The flux densities of the host galaxy of W2246, its AGN component and W2246-NE are $5.2\pm0.2~\mu$Jy, $0.9\pm0.2~\mu$Jy and $6.9\pm0.1~\mu$Jy, respectively. The $K_S$ band photometry of W2246 is taken from \citet{assef2015}, which was observed by Hale P200 WIRC. 

\herschel flux densities and their associated uncertainties have been derived from PSF fitting using \wise 12 \um sources as prior positions. During our fittings, the 12 \um prior positions are fixed, as for the longest passbands of SPIRE (i.e., 350 \um and 500 \um), the increasingly large PSFs make the source of interest and its close neighbor strongly blended. For PACS 70 \um and 160 \um observations which have better spatial resolution, we find little difference in flux measurements if the prior positions are allowed to vary.  Dust continuum emission of W2246 at $\sim 880~$\um has been resolved by ALMA observations and its flux density is $7.4 \pm 0.6~$mJy \citep{diaz-santos2016}. We summarize the photometry of W2246 and W2246f in Table \ref{tab:phot}.

\begin{table*}
\caption{Photometry of W2246 and W2246f}             % title of Table
\label{tab:phot}      % is used to refer this table in the text
\centering                          % used for centering table
\begin{tabular}{lccc}       
%\hline\hline                 % inserts double horizontal lines
\toprule
  Band  				& PSF FWHM [arcsec] & \multicolumn{2}{c}{Flux [mJy]} \\
\cmidrule(l){3-4}
        				&          		& W2246 			& W2246f 			\\    % table heading 
\midrule                       
   SDSS \textit{u}     	& 	1.5 		& -  				& 0.090$\pm$0.006  			\\ 
   SDSS \textit{g}      &	1.4  		& -     			& 0.387$\pm$0.004   	\\
   SDSS \textit{r}      &	1.3	  		& $<3.9^{(a)}$      & 0.752$\pm$0.007  		\\
   SDSS \textit{i}      &	1.2   		& -     			& 1.05$\pm$0.01   		\\
   SDSS \textit{z}      &	1.2	  		& -     			& 1.28$\pm$0.03   		\\ 
   2MASS \textit{J}     &	3.0         & -      			& 1.60$\pm$0.20  		\\
   2MASS \textit{H}     &	3.0         & 0.0052$\pm$0.0002$^{(b)}$  & 1.55$\pm$0.29  	    \\
   2MASS $K_S$        	&	3.0			& 0.0088$\pm$0.0028$^{(c)}$  & 1.43$\pm$0.40   	    \\
   \wise \wone 			&	6.1	 	 	& 0.031$\pm$0.007   & 1.16$\pm$0.01   	   	\\   
   \wise \wtwo   		&	6.4   		& 0.034$\pm$0.007   & 0.70$\pm$0.02  	  	\\
   \wise \wthree   		&	6.5         & 1.2$\pm$0.2      	& 2.52$\pm$0.18     	\\
   \wise \wfour  		&	12.0        & 13.0$\pm$1.6      & 3.17$\pm$0.76         \\
   PACS 70\um       	&	6.0  		& 25.0$\pm$1.3     	& 12.3$\pm$1.2	    	\\ 
   PACS 160\um	        &	11.0   		& 56.1$\pm$2.1  	& 65.8$\pm$2.2          \\ 
   SPIRE 250\um     	&	18.0   		& 75.4$\pm$8.3	    & 70.4$\pm$8.3    	    \\ 
   SPIRE 350\um     	&	25.0   		& 66.0$\pm$8.2		& 29.1$\pm$8.1    	     \\ 
   SPIRE 500\um     	&	38.0   		& 57.0$\pm$13.0		& $<26.0$				  \\ 
   ALMA 880\um      	&	0.4   		& 7.4$\pm$0.6$^{(d)}$      	& -                 \\ 
\bottomrule                          
\end{tabular}
\parbox{160mm} {
\textbf{Notes.} \\
 (a): \citet{tsai2015}; (b) \hst F160W filter with a PSF FWHM of $0.18''$; (c) Hale 200-inch WIRC $K_S$ \citep{assef2015}; (d): \citet{diaz-santos2016}. }
\end{table*}

\section{SED fitting} \label{sec:sedfitting}

\begin{figure}
\centering
\includegraphics[width=\hsize]{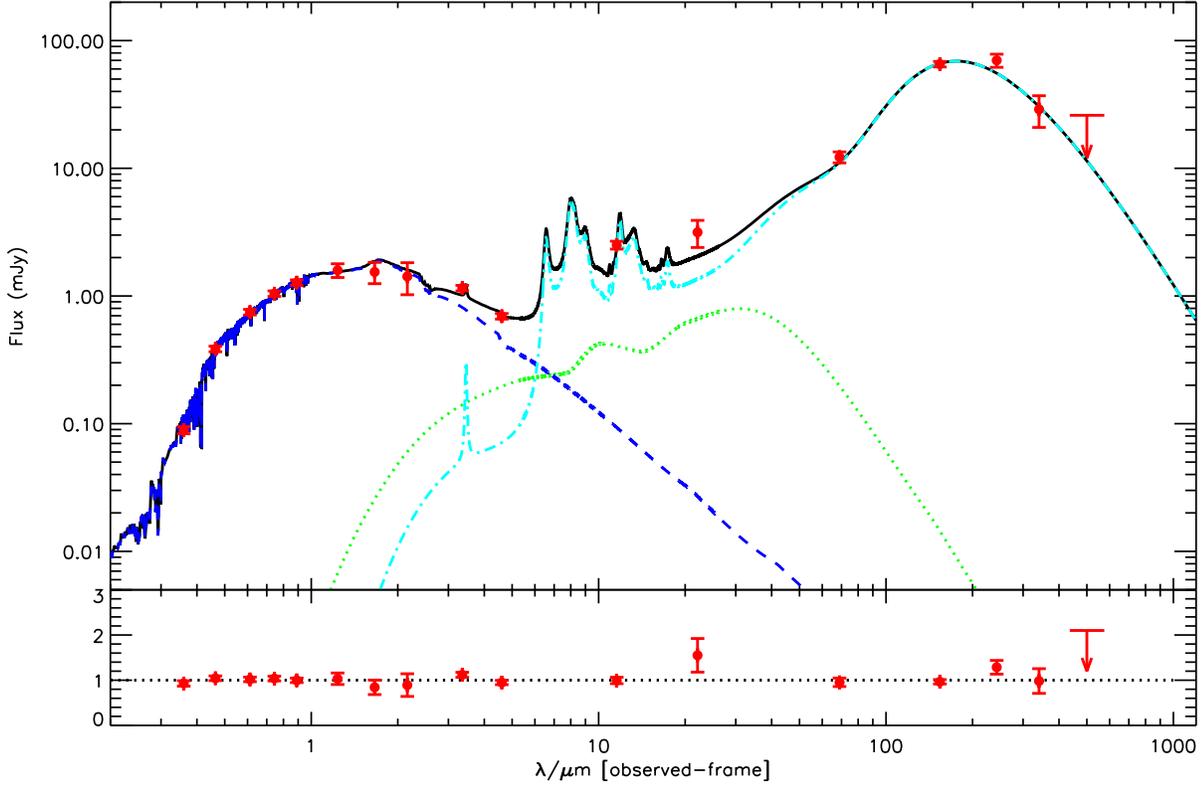}
\caption{Best-fit model SED (solid line) of W2246f with {\sf SED3FIT}. The observed data are listed in Table~\ref{tab:phot}, and plotted with red circles. The blue dashed line represents the attenuated stellar emission and the cyan dot-dashed line represents dust emission from star formation. The green dotted line shows the AGN torus dust emission. The bottom panel shows the ratio of the observed flux to model prediction.}
\label{fig:sedw2246f}
\end{figure}
\begin{figure}
\centering
\includegraphics[width=\hsize]{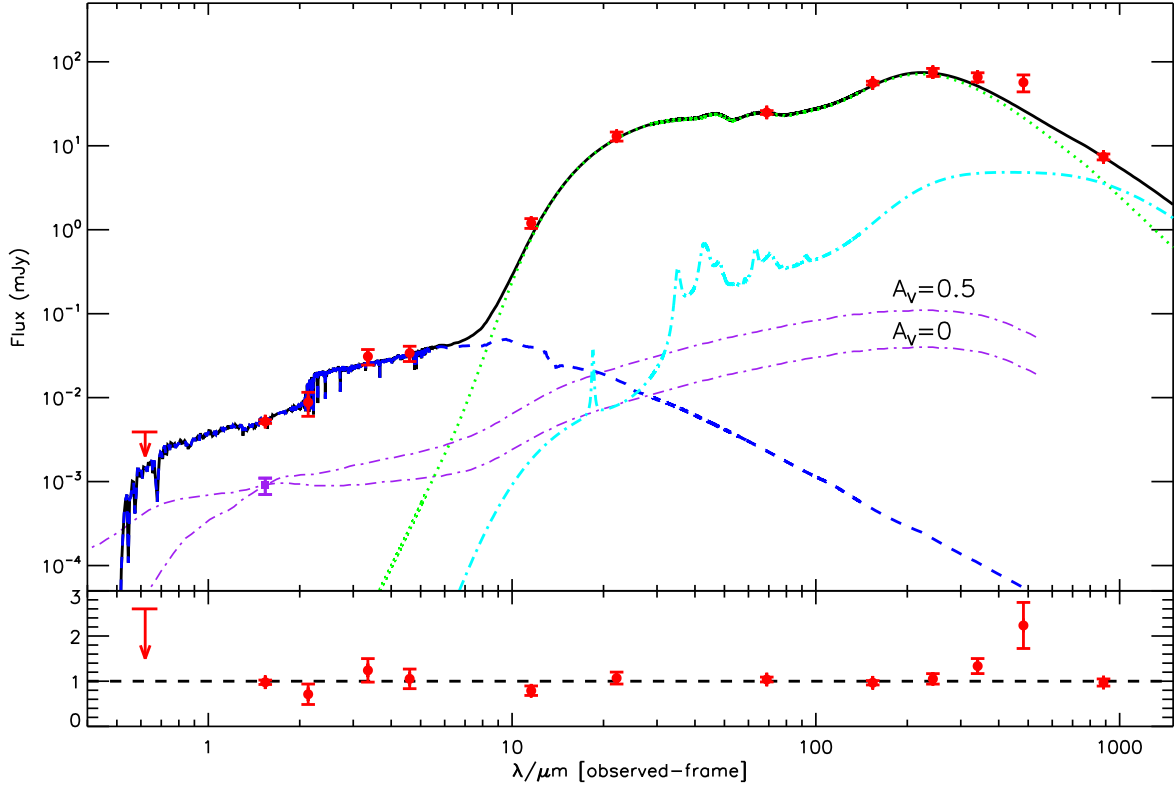}
\caption{Best-fit model SED (solid line) of W2246 with {\sf SED3FIT}. Red circles are the observed data points. The blue dashed and cyan dot-dashed lines show the attenuated stellar emission and dust emission from star formation, respectively. Green dotted line is the contribution of AGN torus emission. The flux contributed by AGN emission in H-band, which is shown as purple filled square, has been estimated based on the structural decomposition using {\sf GALFIT}. We plot the mean SEDs of Type 1 QSOs \citep{richards2006}, unattenuated and attenuated by dust assuming a SMC-like extinction law \citep{prevot1984} with $A_V=0.5$. The unattenuated and attenuated Type 1 QSO SEDs (purple dot-dashed lines) have been normalized to the flux of AGN emission in H-band. The bottom panel shows the ratio of the observed flux to model prediction. }
\label{fig:sedw2246}
\end{figure}
\begin{figure}
\centering
\includegraphics[width=0.6\hsize]{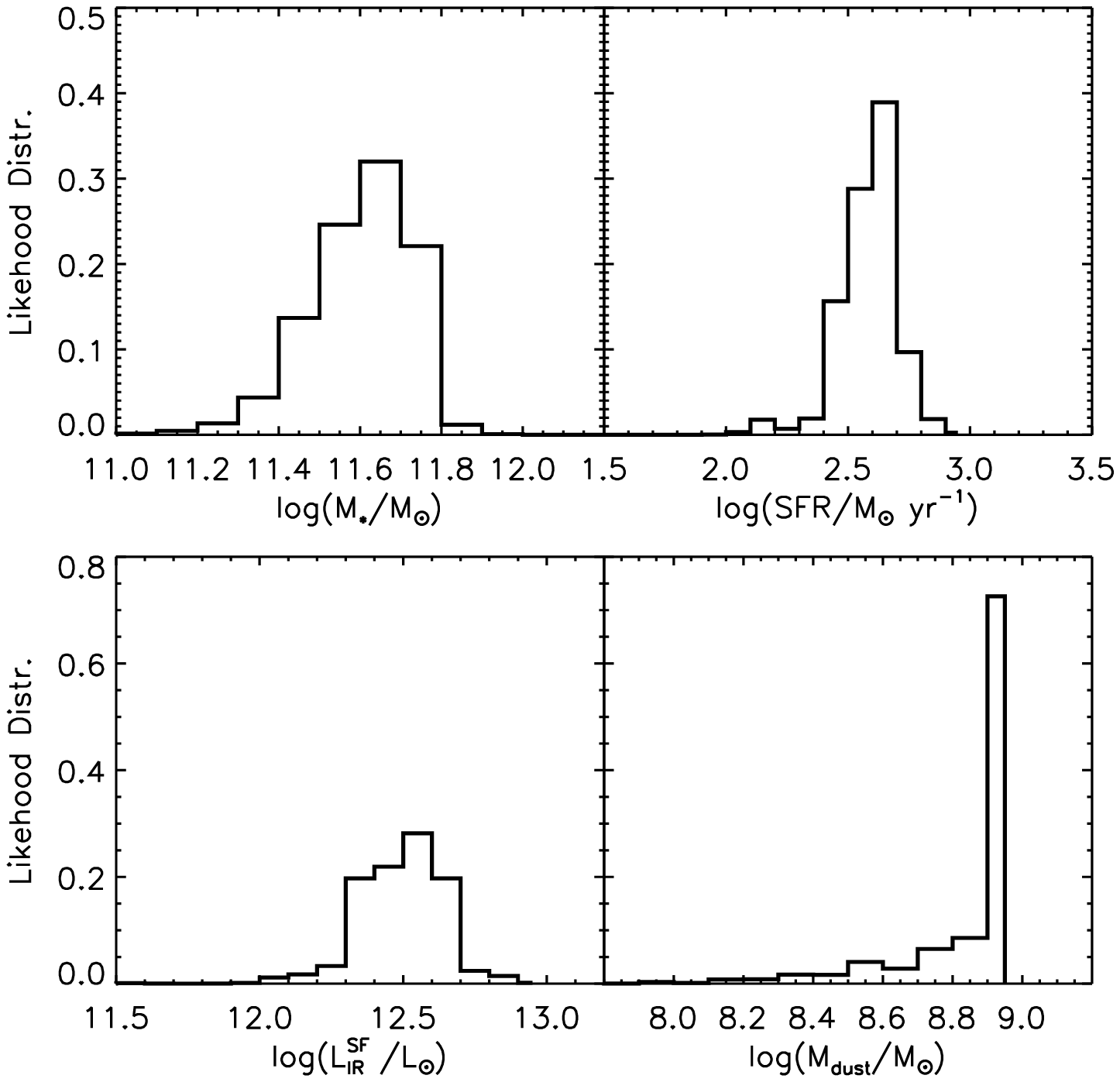}
\caption{Likelihood distributions of stellar mass ($M_\star$), star formation rate (SFR), dust luminosity related to star formation ($L_{\rm IR}^{\rm SF}$) and dust mass ($M_{\rm dust}$) of W2246.       
         }
\label{fig:histw2246}
\end{figure}
%-------

For Hot DOGs, \citet{assef2015} modeled their rest-frame optical through mid-IR SEDs following the approach applied in \citet{eisenhardt2012}. Each SED had been modeled as a combination of the host galaxy template and one AGN SED template. Their result showed that the median value of AGN obscuration for Hot DOGs is $E(B-V)=6.0$. The stellar component dominated the optical-to-near-IR SED, while the AGN component dominated the mid-IR band. In our previous work \citep{fan2016b}, we modeled the IR SEDs of twenty-two Hot DOGs using two main components: dust emission from star formation and AGN torus emission. Our result showed that the two-component model can fit the observed IR SEDs of Hot DOGs well and AGN torus emission dominated the IR energy output.

We construct the rest-frame UV/optical-to-far-IR SED of W2246 using the data set described in Section \ref{sec:data}. At least three components, including stellar emission, dust emission from star formation and AGN torus emission, can contribute to the UV/optical-to-far-IR SED of W2246. Here, we use the three-component SED-fitting code \sedfit by \citet{berta2013}\footnote{http://cosmos.astro.caltech.edu/page/other-tools}, which implements the Multiwavelength Analysis of Galaxy Physical Properties code {\sf MAGPHYS} \citep{dacunha2008}\footnote{http://www.iap.fr/magphys/magphys/MAGPHYS.html} with an additional AGN torus component from the library of \citet{fritz2006} and \citet{feltre2012}, to model the observed SED of W2246. We adopt \citet{bc03} optical/near-IR stellar library and \citet{chabrier2003} initial mass function (IMF). However, we cannot rule out the possibility that a small fraction of the rest-frame UV/optical emission of the luminous AGN is leaked out of the high-obscuration region. As mentioned by \citet{assef2016}, a fraction of the AGN light can be possibly scattered off into our line of sight. We will discuss this possibility in Section \ref{sec:results}.

We use the same three-component \sedfit code to fit the rest-frame UV/optical-to-far-IR SED of W2246f. In Figure \ref{fig:sedw2246f}, solid line shows the best-fit result of W2246f using {\sf SED3FIT}. The dashed and dot-dashed lines represent the attenuated stellar emission and dust emission from star formation, respectively. The dotted line shows the AGN torus emission.

 \section{Results and Discussion} \label{sec:results}

\begin{table}
\caption{Physical properties of W2246 and W2246f}             % title of Table
\label{tab:prop}      % is used to refer this table in the text
\centering                          % used for centering table
\begin{tabular}{lll}       
\hline
\hline                 % inserts double horizontal lines
           & W2246   & W2246f \\ 
\hline
  $L_{\rm IR[8-1000~\mu m]}^{\rm tot}$ [  $L_\odot$ ]       &  $1.2\times10^{14}$   			  &  $3.2\times10^{9}$ \\
  $L_{\rm  IR[8-1000~\mu m]}^{\rm torus}$ [  $L_\odot$ ]    &  $1.1\times10^{14}$   			  &  $2.1\times10^{8}$ \\
  $L_{\rm IR[8-1000~\mu m]}^{\rm SF}$ [  $L_\odot$ ]        &  $4.5\times10^{12}$    			  &  $3.0\times10^{9}$ \\
  $L_{\rm FIR[42-122~\mu m]}^{\rm SF}$ [  $L_\odot$ ]       &  $2.1\times10^{12}$   			  &  $1.0\times10^{9}$ \\
  {\it SFR} [$M_\odot~ yr^{-1}$]                            &  $480^{+70}_{-217}$                 &  $0.09^{+0.07}_{-0.02}$           \\
  $M_\star$ [$M_\odot$]                                     &  $4.3^{+2.3}_{-1.5}\times10^{11}$   &  $1.1^{+0.5}_{-0.2}\times10^{10}$ \\
  $M_{\rm dust}$ [$M_\odot$]                                &  $9.1^{+0.4}_{-4.5}\times10^{8}$    &  $1.4^{+1.6}_{-0.4}\times10^{7}$    \\   
\hline                            
\end{tabular}
\end{table}

In Figure \ref{fig:sedw2246}, we present the best-fit model SED (solid line) of W2246 with {\sf SED3FIT}. The three-component model provides a rather good description of the rest-frame UV/optical-to-far-IR SED with $\chi^2=1.53$. The deviation at 500 \um band is possibly due to the large uncertainty of PSF fitting photometry. Derived physical properties of W2246 have been listed in Table \ref{tab:prop}, including the total IR luminosity ($L_{\rm IR}^{\rm tot}$), the IR luminosity contributed by AGN torus ($L_{\rm IR}^{\rm torus}$), the IR luminosity related to star formation ($L_{\rm IR}^{\rm SF}$), the far-IR luminosity related to star formation ($L_{\rm FIR}^{\rm SF}$ ), star formation rate (SFR), stellar masses ($M_\star$) and dust mass ($M_{\rm dust}$). We plot the likelihood distributions of $M_\star$, SFR, $L_{\rm IR}^{\rm SF}$ and $M_{\rm dust}$ in Figure \ref{fig:histw2246}.

The total IR luminosity is lower than the previous estimations \citep{jones2014,fan2016b} by a factor of about 2 due to taking out the contamination of the foreground galaxy W2246f to SPIRE photometry. AGN torus emission, contributing over 95\% of the total IR luminosity, not only dominates in the mid-IR wavelength range, but also has a significant contribution up to the rest-frame $100~ \mu m$. At the rest-frame wavelength range longer than $100~ \mu m$, dust emission related to star formation starts to dominate. After deducting the AGN contribution, the far-IR luminosity related to star formation is only $2.1\times10^{12}$ $L_\odot$, which is lower than that used in \citet{diaz-santos2016} by one order of magnitude. Adopting the total [CII] luminosity of W2246 $L_{\rm [CII]} = 6.1 \times 10^9$ $L_\odot$, our estimation of the far-IR luminosity results in the [CII]-to-far-IR emission ratio ([CII]/FIR) of $2.9 \times 10^{-3}$, which is similar to some high-redshift ULIRGs \citep{delooze2014} and quasars at $z>4$ \citep{willott2013,willott2015,venemans2016}. The previously reported [CII]/FIR deficit of W2246 \citep{diaz-santos2016} and other high-redshift quasars \citep{wang2013} may be at least in part due to AGN contamination of the far-IR emission. 

UV/Optical SED is dominated by stellar emission. The derived stellar mass of W2246 is $4.3\times10^{11}~M_\odot$, which is among the most massive galaxies with spectroscopic redshift $z>4.5$ \citep{caputi2015}. In order to consider the effects of different IMF, metallicity and star formation history (SFH) on the stellar mass estimation, we utilize {\it FAST} \citep{kriek2009} to fit the observed $H$, $K_S$, \wone and \wtwo bands. The derived stellar mass can change by a factor of up to 0.3 dex, adopting the different combination of IMF, metallicity and SFH. It is possible that the stellar mass can be overestimated by the contamination of AGN emission (for instance, the scattered AGN emission) to optical/near-IR bands. We consider the possible contribution of AGN emission by doing the structural decomposition using the \galfit package \citep{peng2002,peng2010}. The high spatial resolution \hst H-band image of W2246 has been decomposed with a \ser+ PSF  model. The decomposed PSF component has the flux $0.9\pm0.2~\mu Jy$,  which is about six times weaker than the \ser component. We assume that the PSF component comes from the scattered AGN emission as suggested by \citet{assef2016}. In Figure \ref{fig:sedw2246}, an attenuated Type 1 QSO SED \citep{richards2006} has been plotted to present this scattered AGN emission. The dust attenuation of host galaxy has been set to $A_V=0.5$, which is determined by the \sedfit result. The scattered AGN emission is lower than stellar emission by over one order of magnitude in the optical and near-IR bands. Thus, the AGN contamination has the negligible effect on the stellar mass measurement of W2246. 

The structural parameters, \ser index $n = 0.7$ and effective radius $R_e = 1.3$ Kpc, of W2246 have been derived based on the \ser+ PSF model, making it be a disk-like, extremely compact galaxy at such a high redshift.  \citet{diaz-santos2016} showed even more compact structures of  [CII] emission line and dust continuum than UV continuum. Such a compact galaxy is expected to evolve into a red nugget at $z\sim 2-3$ and experience a dramatic structural evolution. In order to catch up the local mass-size relation of massive early-type galaxies \citep{shen2003}, W2246 requires to increase its present size by a factor of $\sim 7$. AGN feedback, which is taking action to blow out the ISM in W2246 \citep{diaz-santos2016}, will possibly plays an important role in such a dramatic size increase, as suggested by our previous model \citep{fan2008,fan2010}. Another mechanism like dry minor merger \citep[e.g.,][]{naab2009} may also contribute the observed size evolution during the late evolutionary stage. Recently, a new scenario to explain the evolution of extremely compact galaxies at high redshift supposes that they survive as the compact cores (bulge components) embedded in present-day massive galaxies \citep{graham2015,delarosa2016}. 

The derived SFR of W2246 is 480 $M_\odot~yr^{-1}$, which is comparable to some starburst galaxies. However, considering its high redshift and large stellar mass, W2246 still lies below the the star forming galaxy (SFG) main sequence (MS) which suggests SFR$\sim1100~M_\odot~yr^{-1}$ for a MS galaxy with $4.3\times10^{11}~M_\odot$ stellar mass and at the age of the universe $t\sim1.2$ Gyr  \citep{speagle2014}. This result suggests that W2246 may be experiencing the declining and quenching of star formation. 

By integrating the best-fit model SED of W2246, we derive its bolometric luminosity $L_{\rm bol} = 1.7 \times 10^{14}~L_\odot$. Assuming that the SMBH in the center of W2246 accretes at the Eddington ratio $\eta = 1$ \citep{wu2018},  the estimated black hole mass is $5.1 \times 10^9~M_\odot$. The corresponding black hole-bulge mass ratio ($M_{\rm BH}/M_{\rm bulge}$) is 0.012 which is about 2.4 times higher than the present-day value, suggesting  that the SMBH accumulates most of its mass before the formation of the stellar bulge. Both $M_{\rm BH}$ and $M_{\rm BH}/M_{\rm bulge}$ ratio of W2246 are in agreement with those of many other high-redshift quasars \citep[e.g.,][]{peng2006,wang2010}. The present-day $M_{\rm BH}/M_{\rm bulge}$ ratio, which has been recently updated by \citet{kormendy2013}, is about 0.0049 at $M_{\rm bulge} = 10^{11} M_\odot$ and is 2-4 times larger than previous values ranging from 0.001 to 0.0023.  Considering the large intrinsic scatter (0.29 dex) of the present-day $M_{\rm BH}/M_{\rm bulge}$ ratio, only a moderate evolution of the BH mass ratio of W2246 is required to reach the present-day $M_{\rm BH}-M_{\rm bulge}$ relation. W2246 is expected to evolve towards the most massive galaxy hosting monster black hole in the local Universe. 
%For instance, NGC~3842, one of the nearby Brightest Cluster Galaxies, shows similar values of $M_{\rm BH}$ and $M_{\rm bulge}$ \citep{mcconnell2012}.
 
The derived dust mass of $M_{\rm dust} = 9.1\times10^{8}~M_\odot$ indicates that there is likely a large amount of molecular gas ($\sim 10^{11}~M_\odot$)  in W2246. The idea that luminous Hot DOGs may have plenty of molecular gas has been supported by our recent ALMA CO observations of three Hot DOGs \citep{fan2017b}, which find that all of them have a significant molecular gas reservoir ($\sim10^{10-11} M_\odot$). The ongoing ALMA CO observations of W2246 will help measure its molecular gas directly. 
 
\section{Conclusions} \label{sec:conc}

W2246, a {\it WISE}-selected, hyperluminous dust-obscured galaxy at $z=4.593$, was taken as the most luminous galaxy known in the Universe. However, according to the multiwavelength images (see Figure \ref{fig:stamps}), we noted that the previous \herschel SPIRE photometry of W2246 was contaminated by a foreground galaxy (W2246f),  resulting in an overestimation of its total IR luminosity. Based on the new \wise and \herschel SPIRE photometry, we perform a SED analysis on the rest-frame UV/optical-to-far-IR of W2246 with {\sf{SED3FIT}}. The derived total IR luminosity is about 2 times lower than the previous estimations, making it be not the most luminous Hot DOG any more. 

The results from the new SED-fitting show that W2246 is a very interesting object, being in a key transition phase during the evolution of massive galaxies. With the derived stellar mass $M_\star = 4.3\times10^{11}~M_\odot$, it is among the most massive galaxies with spectroscopic redshift $z>4.5$. Besides the high stellar mass, its structure is extremely compact, which indicates that it will experience a dramatic size evolution towards low redshift. Most of ($>95\%$) its IR luminosity is from AGN torus emission, revealing the rapid growth of the central SMBH according to the accretion. 
%The estimated $M_{\rm BH}$ and $M_{\rm BH}/M_{\rm bulge}$ ratio are similar to those of the Brightest Cluster Galaxies in the local Universe. 
Although the derived SFR is high (480 $M_\odot~yr^{-1}$), it still lies below the star-forming galaxy main sequence. Therefore, it has been suggested to be experiencing the declining and quenching of star formation. We also predict that W2246 may have a significant molecular gas reservoir, which can be tested by the ongoing ALMA CO line observations.
Both AGN and star formation activities in W2246 may be related to its environment. It is possible that W2246 lies in an overdense environment, which has been suggested by several previous works \citep[e.g.,][]{jones2014,assef2015,fan2017a}. Our ongoing work with the VLT FORS2 narrow-band imaging will shed insight on the environment of Lyman-alpha emitters (LAEs) around it.

\begin{acknowledgements}

The authors would like to thank the anonymous referee for his/her comments and suggestions, which have greatly improved this paper. We thank Dr. Chen Cao and Ms. Berzaf Berhane for their valuable discussions on Herschel photometry and the usage of {\sf SED3FIT}. LF acknowledges the support from National Key Research and Development Program of China (No. 2017YFA0402703). This work is supported by the National Natural Science Foundation of China (NSFC, Grant Nos. 11773020, 11433005 and 11573001) and Shandong Provincial Natural Science Foundation, China (ZR2017QA001). KK acknowledges the Knut and Alice Wallenberg Foundation for support. 

\facility{\textit{WISE}, \textit{Herschel} (PACS,SPIRE), \textit{HST}, \textit{SDSS} }.

\end{acknowledgements}

%% This command is needed to show the entire author+affilation list when
%% the collaboration and author truncation commands are used.  It has to
%% go at the end of the manuscript.
%\allauthors

%% Include this line if you are using the \added, \replaced, \deleted
%% commands to see a summary list of all changes at the end of the article.
%\listofchanges


\begin{thebibliography}{99}

\bibitem[Assef et al.(2015)]{assef2015} Assef, R.~J., Eisenhardt, P.~R.~M., Stern, D., et al.\ 2015, \apj, 804, 27
\bibitem[Assef et al.(2016)]{assef2016} Assef, R.~J., Walton, D.~J., Brightman, M., et al.\ 2016, \apj, 819, 111
\bibitem[Berta et al.(2013)]{berta2013} Berta, S., Lutz, D., Santini, P., et al.\ 2013, \aap, 551, A100 
\bibitem[Bridge et al.(2013)]{bridge2013} Bridge, C.~R., Blain, A., Borys, C.~J.~K., et al.\ 2013, \apj, 769, 91
\bibitem[Bruzual \& Charlot(2003)]{bc03} Bruzual, G., \& Charlot, S.\ 2003, \mnras, 344, 1000 
\bibitem[Caputi et al.(2015)]{caputi2015} Caputi, K.~I., Ilbert, O., Laigle, C., et al.\ 2015, \apj, 810, 73 
\bibitem[Chabrier(2003)]{chabrier2003} Chabrier, G.\ 2003, \pasp, 115, 763 
\bibitem[Cutri et al.(2013)]{cutri2013} Cutri, R.~M., \& et al.\ 2013, VizieR Online Data Catalog, 2328, 0
\bibitem[da Cunha et al.(2008)]{dacunha2008} da Cunha, E., Charlot, S., \& Elbaz, D.\ 2008, \mnras, 388, 1595
\bibitem[D{\'{\i}}az-Santos et al.(2016)]{diaz-santos2016} D{\'{\i}}az-Santos, T., Assef, R.~J., Blain, A.~W., et al.\ 2016, \apjl, 816, L6
\bibitem[de la Rosa et al.(2016)]{delarosa2016} de la Rosa, I.~G., La Barbera, F., Ferreras, I., et al.\ 2016, \mnras, 457, 1916 
\bibitem[De Looze et al.(2014)]{delooze2014} De Looze, I., Cormier, D., Lebouteiller, V., et al.\ 2014, \aap, 568, A62
\bibitem[Eisenhardt et al.(2012)]{eisenhardt2012} Eisenhardt, P.~R.~M., Wu, J., Tsai, C.-W., et al.\ 2012, \apj, 755, 173
\bibitem[Fan et al. (2008)]{fan2008} Fan, L., Lapi, A., De Zotti, G., \& Danese, L.  2008, \apjl,689, L101
\bibitem[Fan et al. (2010)]{fan2010} Fan, L., Lapi, A., Bressan, A., et al.  2010, \apj, 718, 1460
\bibitem[Fan et al.(2016a)]{fan2016a} Fan, L., Han, Y., Fang, G., et al.\ 2016, \apjl, 822, L32 
\bibitem[Fan et al.(2016b)]{fan2016b} Fan, L., Han, Y., Nikutta, R., Drouart, G., \& Knudsen, K.~K.\ 2016, \apj, 823, 107
\bibitem[Fan et al.(2017a)]{fan2017a} Fan, L., Jones, S.~F., Han, Y., \& Knudsen, K.~K.\ 2017, \pasp, 129, 124101 
\bibitem[Fan et al.(2017b)]{fan2017b} Fan, L., Knudsen, K.~K., Fogasy, J., \& Drouart, G.\ 2017, arXiv:1711.10615 
\bibitem[Feltre et al.(2012)]{feltre2012} Feltre, A., Hatziminaoglou, E., Fritz, J., \& Franceschini, A.\ 2012, \mnras, 426, 120
\bibitem[Fritz et al.(2006)]{fritz2006} Fritz, J., Franceschini, A., \& Hatziminaoglou, E.\ 2006, \mnras, 366, 767 
\bibitem[Graham et al.(2015)]{graham2015} Graham, A.~W., Dullo, B.~T., \& Savorgnan, G.~A.~D.\ 2015, \apj, 804, 32 
\bibitem[Griffin et al.(2010)]{griffin2010} Griffin, M.~J., Abergel, A., Abreu, A., et al.\ 2010, \aap, 518, L3
\bibitem[Jones et al.(2014)]{jones2014} Jones, S.~F., Blain, A.~W., Stern, D., et al.\ 2014, \mnras, 443, 146
\bibitem[Komatsu et al.(2011)]{komatsu2011} Komatsu, E., Smith, K.~M., Dunkley, J., et al.\ 2011, \apjs, 192, 18 
\bibitem[Kormendy \& Ho(2013)]{kormendy2013} Kormendy, J., \& Ho, L.~C.\ 2013, \araa, 51, 511 
\bibitem[Kriek et al.(2009)]{kriek2009} Kriek, M., van Dokkum, P.~G., Labb{\'e}, I., et al.\ 2009, \apj, 700, 221
\bibitem[Krist et al.(2011)]{krist2011} Krist, J.~E., Hook, R.~N., \& Stoehr, F.\ 2011, \procspie, 8127, 81270J
\bibitem[Lang(2014)]{lang2014} Lang, D.\ 2014, \aj, 147, 108 
%\bibitem[McConnell et al.(2012)]{mcconnell2012} McConnell, N.~J., Ma, C.-P., Murphy, J.~D., et al.\ 2012, \apj, 756, 179 
\bibitem[Naab et al.(2009)]{naab2009} Naab, T., Johansson, P.~H., \& Ostriker, J.~P.\ 2009, \apjl, 699, L178 
\bibitem[Peng et al.(2002)]{peng2002} Peng, C.~Y., Ho, L.~C., Impey, C.~D., \& Rix, H.-W.\ 2002, \aj, 124, 266
\bibitem[Peng et al.(2006)]{peng2006} Peng, C.~Y., Impey, C.~D., Rix, H.-W., et al.\ 2006, \apj, 649, 616
\bibitem[Peng et al.(2010)]{peng2010} Peng, C.~Y., Ho, L.~C., Impey, C.~D., \& Rix, H.-W.\ 2010, \aj, 139, 2097 
%\bibitem[Pilbratt et al.(2010)]{pilbratt2010} Pilbratt, G.~L., Riedinger, J.~R., Passvogel, T., et al.\ 2010, \aap, 518, L1
\bibitem[Poglitsch et al.(2010)]{poglitsch2010} Poglitsch, A., Waelkens, C., Geis, N., et al.\ 2010, \aap, 518, L2
\bibitem[Prevot et al.(1984)]{prevot1984} Prevot, M.~L., Lequeux, J., Prevot, L., Maurice, E., \& Rocca-Volmerange, B.\ 1984, \aap, 132, 389 
\bibitem[Richards et al.(2006)]{richards2006} Richards, G.~T., Lacy, M., Storrie-Lombardi, L.~J., et al.\ 2006, \apjs, 166, 470
\bibitem[Shen et al. (2003)]{shen2003} Shen, S., Mo, H.~J.,White, S.~D.~M., et al.\ 2003, \mnras, 343, 978
\bibitem[Speagle et al.(2014)]{speagle2014} Speagle, J.~S., Steinhardt, C.~L., Capak, P.~L., \& Silverman, J.~D.\ 2014, \apjs, 214, 15
\bibitem[Tsai et al.(2015)]{tsai2015} Tsai, C.-W., Eisenhardt, P.~R.~M., Wu, J., et al.\ 2015, \apj, 805, 90
\bibitem[Venemans et al.(2016)]{venemans2016} Venemans, B.~P., Walter, F., Zschaechner, L., et al.\ 2016, \apj, 816, 37 
\bibitem[Wang et al.(2010)]{wang2010} Wang, R., Carilli, C.~L., Neri, R., et al.\ 2010, \apj, 714, 699 
\bibitem[e.g., Wang et al.(2013)]{wang2013} Wang, R., Wagg, J., Carilli, C.~L., et al.\ 2013, \apj, 773, 44 
\bibitem[Willott et al.(2013)]{willott2013} Willott, C.~J., Omont, A., \& Bergeron, J.\ 2013, \apj, 770, 13 
\bibitem[Willott et al.(2015)]{willott2015} Willott, C.~J., Bergeron, J., \& Omont, A.\ 2015, \apj, 801, 123 
\bibitem[Wright et al.(2010)]{wright2010} Wright, E.~L., Eisenhardt, P.~R.~M., Mainzer, A.~K., et al.\ 2010, \aj, 140, 1868
\bibitem[Wu et al.(2012)]{wu2012} Wu, J., Tsai, C.-W., Sayers, J., et al.\ 2012, \apj, 756, 96
\bibitem[Wu et al.(2018)]{wu2018} Wu, J., Jun, H.~D., Assef, R.~J., et al.\ 2018, \apj, 852, 96 

\end{thebibliography}
\end{document}